A Study on Altering PostgreSQL from Multi-Processes Structure to Multi-Threads Structure

Zhiyong Shan

Department of Computer Science, Renmin University of China

**This Paper is in Chinese**

**Abstract**: How to altering PostgreSQL database from multi-processes structure to multi-threads structure is a difficult problem. In the paper, we bring forward a comprehensive alteration scheme. Especially, put rational methods to account for three difficult points: semaphores, signal processing and global variables. At last, applied the scheme successfully to modify a famous open source DBMS.

**Keywords**: Database, Multi-Threads, Operating System

## 1．Introduction

PostgreSQL 是著名的开放源代码数据库，对数据库发展有着重要影响，同时具有广泛的研究和应用价值。它起源于斯坦福大学，已经持续开发了二十多年。目前技术比较先进的商业数据库一般采用多线程结构，以提高效率，但是，由于 PostgreSQL 的源码有 25 万行之多，而且历史沉积时间长，改造的难度比较大，所以一直保留多进程的结构，国际上还没有人成功将 linux 上的 PostgreSQL 改造为多线程结构[9]。

很多的其他大型软件也是基于多进程结构开发的，如果将它们改造成多线程结构，可以显著地提高性能[1][2][3][4]。

多线程结构性能提高的原因是线程之间共享相同的进程空间，线程启动和通信所消耗的系统资源比进程少得多。如表 1 数据[1]：

表 1

| 操作 | 花费时间（微秒） |
| --- | --- |
| 创建线程 | 52 |
| 创建进程 | 1700 |
| 线程间进行 PV 原语 | 66 |
| 进程间进行 PV 原语 | 200 |

数据表明，创建一个新进程所花时间大概是创建一个新线程的 30 倍，通信开销是 3 倍。

对于大型软件，源代码数量浩大，往往超过 10 万行，而多线程改造时需要涉及的代码散

布在各处，如果没有正确全面的方法，很难改造成功。本文系统全面地阐述在 Linux 平台上 PostgreSQL 多线程改造时的技术方案，具有较好的参考价值。

文中，首先分析 PostgreSQL 的多进程结构，然后提出详细的多线程改造方案，再指出调试过程中发现的 PostgreSQL 的一个小的设计缺陷，最后进行性能测试。

## 2．PostgreSQL 的多进程结构分析

### 2．1 总体描述

在支持多用户的情况下（以后的讨论均针对多用户的情况），PostgreSQL 后台由以下几种进程组成：

表 2

|    | 进程名 | 功能 |
|----|------|------|
| 1. | Postmaster 进程 | 是一个后台的守护进程，负责启动其他进程，初始化系统，循环监听客户端的请求并分发给 Postgres 进程处理。 |
| 2. | Postgres 进程 | 负责实际处理客户端的数据库请求。 |
| 3. | StartupDataBase 进程 | 启动数据库（根据 sga.scf 文件来做） |
| 4. | ShutdownDataBase 进程 | 关闭数据库 |
| 5. | CheckPointDataBase 进程 | 执行检查点 |
| 6. | Pgstat 父进程 | 缓冲统计信息 |
| 7. | Pgstat 子进程 | 收集统计信息 |

其中，StartupDataBase 进程、ShutdownDataBase 进程和 CheckPointDataBase 进程都是由函数 BootstrapMain()来实现的，所以通称为 Bootstrap 进程。Pgstat 父子进程通称为 Pgstat 进程。

### 2．2 进程间的派生关系

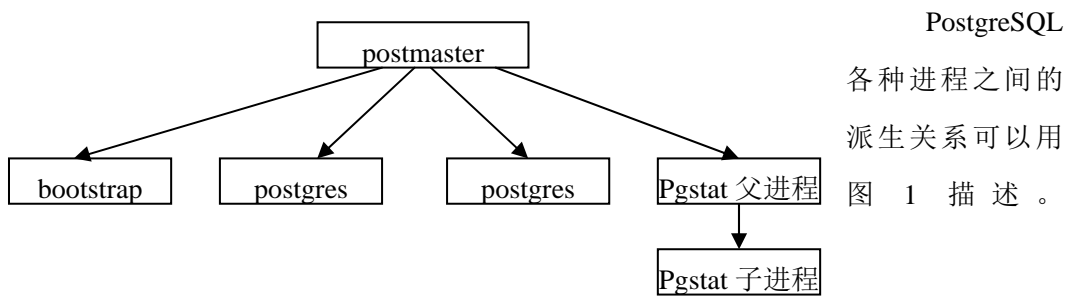

PostgreSQL 各种进程之间的派生关系可以用图 1 描述。

图 1 PostgreSQL 各种进程之间的派生关系

Postmaster 作为守护进程一直运行着，它派生出 postgres、bootstrap、pgstat 父子进程。

## 2．3 进程间的通信关系

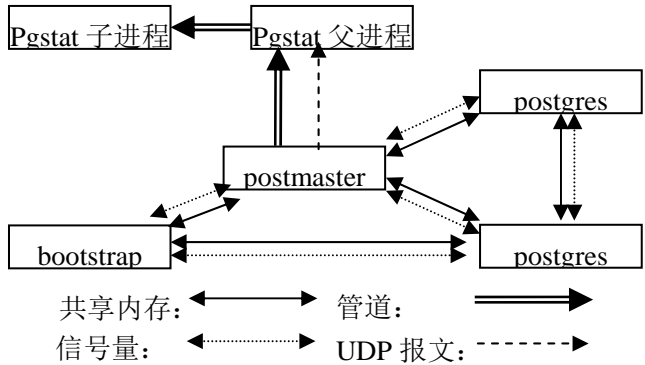

PostgreSQL 各种进程之间的通信关系可以用图 2 描述。Postmaster、postgres、bootstrap 之间使用共享内存和信号量通信，共享内存中保存着上述三种进程所共享的重要数据结构，信号量为各种数据库锁的实现提供底层支持。Pgstat 父子进程之间

图 2 PostgreSQL 各种进程之间的通信

使用管道通信，借助管道 Pgstat 子进程可以感知父进程是否退出。Pgstat 父进程与 postmaster 之间使用管道和 UDP 报文通信，管道用来通知 Pgstat 父进程 postmaster 是否退出，UDP 报文用来发送统计信息。

## 3．改造方案

由于 PostgreSQL 代码量巨大，改造工作很容易引入错误，造成系统不稳定，因此，基本的改造思想是：达到线程改造目的的前提下，尽量少地改动代码。

依据前面的多进程结构分析，Pgstat 父子进程比较独立，它们主要通过 UDP 与 postmaster 通信，对于一个数据库后台守护进程，只需要启动一对 pgstat 父子进程。因此，可以保持其进程结构，而只将 postmaster、postgres 和 bootstrap 改造成线程。

将多进程结构改造为多线程结构，从理论上说，应该涉及以下七个方面的改造[5][6][7]：进程控制、进程间通信（包括：共享内存、信号量和管道）、信号处理、全局变量、进程资源限制、动态内存回收和线程安全。以下从这七个方面阐述改造方案。

## 3．1 进程控制

进程控制[5][6]主要涉及进程的启动、终止、环境变量、会话、进程标识、进程睡眠、父子进程同步等方面。可以用下表描述对进程控制部分的改造：

表 3

| 功能 | 进程特有函数 | 对应的线程函数 | 改造方法 |
|---|---|---|---|
| 启动 | Main(),Fork() | Pthread_create() | 改为启动线程 |
| 终止 | 从 main 函数返回 | 从过程函数返回，返回状态值就是过程函数的返回值。 | 从过程函数返回 |
|  | 调用 exit 函数 | Pthread_exit() | 都改为调用 pthread_exit() |
|  | 调用_exit 函数 | | |
|  | 调用 abort 函数 | | |
| 环境变量 | Getenv() | 没有对应的函数 | 不改造 |
|  | Putenv() | 没有对应的函数 | |
| 会话 | Setsid() | 没有对应的函数 | 由于主要目的是使进程脱离终端，所以不必改造 |
| 进程标识 | Getpid() Getppid() | Pthread_self() | 创建线程时将线程 ID 保存在一个全局数组中 |
| 父子进程同步 | Waitpid() Wait3() Wait4() | 没有对应的函数 | 用 pthread_join()函数和 pthread_detach()函数实现父子线程同步 |
| 进程睡眠 | Sleep() | Pthread_delay_np() | 将所有线程的 sleep()改为 Pthread_delay_np() |

## 3．2 进程间通信

进程间通信主要包括：管道和 SystemV IPC。SystemV IPC 有消息队列、信号量和共享内存三种。

### 3．2．1 共享内存

改造共享内存机制的总体思路是：建立 SystemV 共享内存机制的模拟层，上层程序不作改动。模拟层在内部使用用户空间的内存，而不是核心空间的内存。需要模拟的函数包括：shmget(),shmctl(),shmat(),shmdt()。

需要注意的是，SystemV 共享内存机制的 shmdt()函数给共享内存段的引用计数减1，当引用计数为 0 时，由内核自动删除共享内存。所以，模拟层的 shmdt()函数要在

内部自动释放所分配的内存。

### 3．2．2 信号量

改造信号量机制的总体思路是：建立 SystemV 信号量机制的模拟层，上层程序不作改动。当信号量的值为 1 时，模拟层在内部使用线程的互斥锁模拟信号量。当信号量的值大于 1 时，只有撰写代码来模拟。需要模拟的信号量函数包括：semget(),semctl(),semop()。

实际上，线程的互斥锁与操作系统内核中实现的信号量机制在释放锁（或信号量）的时候有差别：

- 操作系统内核中的信号量机制：释放信号量时，操作系统核心将唤醒等待该信号量的进程。
- 线程的互斥锁机制：释放锁时，不唤醒等待该锁的进程，需要由应用程序自己唤醒。

如果不考虑这个差别，将导致等待互斥锁的线程无限等待。所以，我们进一步采用互斥锁和信号机制相结合的方式模拟操作系统的信号量机制。信号机制的主要作用是在释放互斥锁时唤醒等待线程。这样，释放与获取信号量的函数流程如下：

释放信号量的函数：

*释放互斥锁;*
*发信号给等待线程;*

获取信号量的函数：

*while(尝试加锁并且返回标志为不成功){*
  *设置线程的信号掩码;*
  *if(等待信号并且成功){　//在此处等待信号*
    *continue;　//有人解锁了，再尝试加锁*
  *}*
*};*

### 3．2．3 管道

改造管道机制不能采用模拟层的方式，因为管道机制是依循 UNIX 的泛文件思想建立，其使用的 read()、write()、close()、fcntl()、fstat()、dup()、dup2()等函数，与文件操作函数相同。在多进程的情况下，管道主要用于进程之间的通信，应用于同一进

程是没有意义的。但是，在多线程的情况下，仍然可以沿用，不过，管道机制还应该区分无名管道和命名管道进行一些改造。

无名管道：

需要将各个线程原来代码中的 close()函数调用删去，以避免关闭管道，因为多个线程处于同一进程之中，需要同时使用管道的读端和写端。另外，需要设置管道的 O_NDELAY 标志，避免整个进程被堵塞。

命名管道：

需要在调用 open()函数时，指定 O_NONBLOCK 标志，避免整个进程被堵塞。

## 3．3 信号处理

多线程下的信号处理是一个比较复杂的问题。由于线程机制实现上的缺陷，对于同一个信号，同一进程的所有线程只能共享一个信号处理函数。要将原来的多进程结构的信号处理机制改造为多线程的信号处理机制，关键在于如何在一个所有线程统一的信号处理函数中正确地完成所需要的处理。可以通过以下几个步骤实现：

（1） 按照信号类型建立进程的信号处理函数。进程的信号处理函数实际是分流器，将信号处理任务分发到各个线程的相应信号处理函数。进程的信号处理函数在进程各个线程中完成安装。

（2） 每个信号类型建立一个信号处理分流标记，将这些标记设为线程私有变量。在线程代码中设置信号处理函数的位置同时设置线程的信号处理分流标记的值。信号发送到线程后，进程信号处理函数依据信号处理分流标记，调用相应的信号处理函数（相当于原来进程的信号处

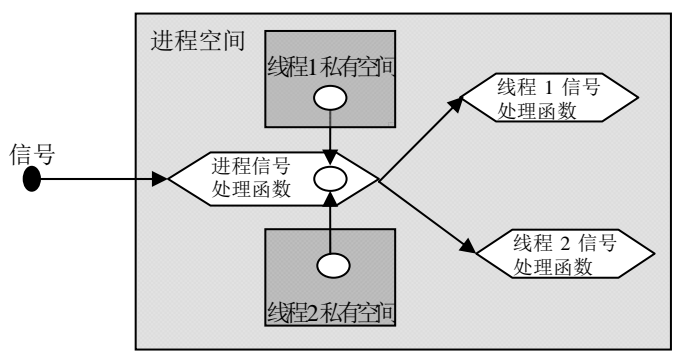

注：白色圆圈为信号处理分流标记及其在各个线程私有空间的拷贝

图 3 多线程下的信号处理机制原理

理函数的代码）完成信号处理。

工作原理如图 3 所示。

## 3．4 全局变量

多线程改造中，全局变量的改造是一个棘手的问题。在多进程结构下，全局变量属于各个进程独有，其他进程无法修改。而在多线程结构下，全局变量由各个线程共享，每个线程都可以读写修改，很容易造成不一致和冲突。而且，由于全局变量散布在整个代码的各个部分，改动的牵涉面比较广。解决这个问题的方案如下：

（1）将所有全局变量和全局静态变量组织成一个大的结构，每个全局变量和全局静态变量是结构的一个成员。对于全局变量，其成员名就是变量名；对于全局静态变量，其成员名是所在文件名加变量名。

（2）定义一个全局的指向结构的指针变量，该指针变量是一个私有变量

（3）在定义全局变量的位置，将全局变量定义成宏，有如下的形式：#define 全局变量名 结构指针->对应的成员

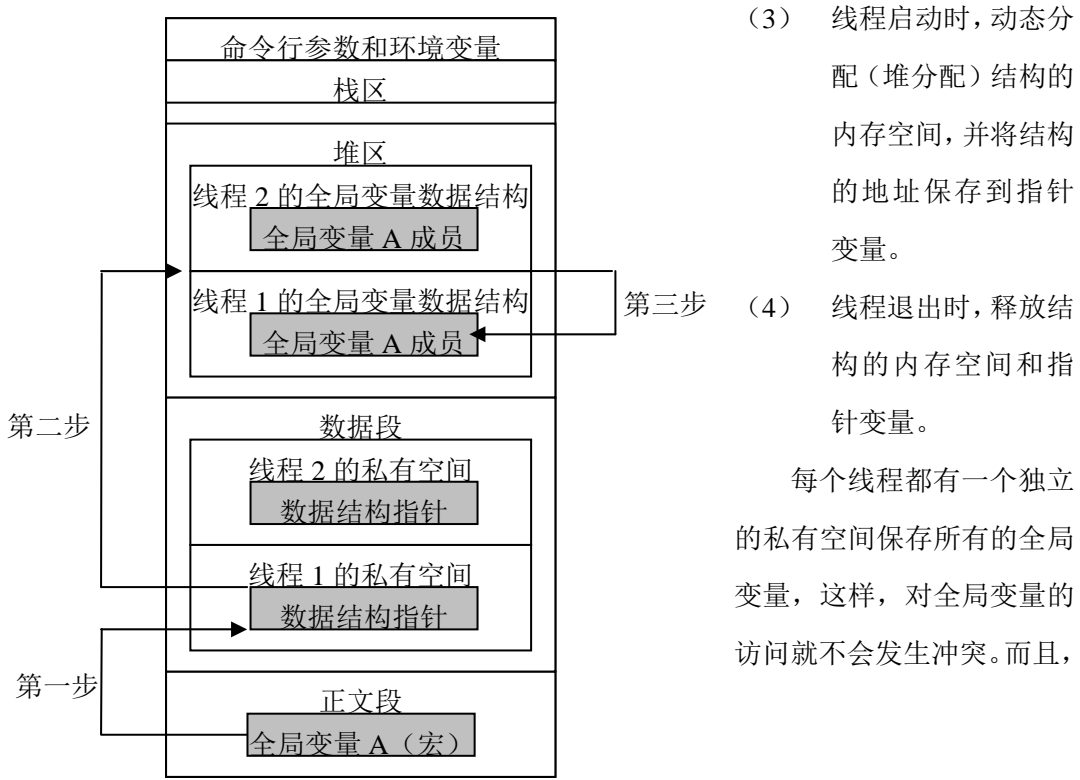

（3）线程启动时，动态分配（堆分配）结构的内存空间，并将结构的地址保存到指针变量。

（4）线程退出时，释放结构的内存空间和指针变量。

每个线程都有一个独立的私有空间保存所有的全局变量，这样，对全局变量的访问就不会发生冲突。而且，

图 4 全局变量实现机制的寻址原理

由于采用与全局变量名相同的宏，所以不必在代码的各处进行改造。全局变量实现机制的寻址原理如图 4。

第一步：调用获取线程私有变量的函数，从线程私有空间得到当前线程的数据结构指针。

第二步：由数据结构指针在堆中找到保存全局变量数据结构的首地址。

第三步：在全局变量数据结构中找到全局变量的值。

## 3．5 进程资源限制

每个进程都有其可用资源限制，特别是每个进程只能打开 1024 个文件。当程序改造为多线程结构时，由于资源限制的原因，将影响程序的运行。Unix 系统规定只有超级用户才能改变进程资源的硬限制，所以我们采用以下的方案解决进程资源限制问题。

- 将程序的可执行文件的属主设为 root,并设定 setuid 位。
- 建立函数 break_resource_limit()，该函数在 main()中被调用，主要是提高进程资源的软硬限制，并在函数结尾将进程的 euid 重新设为普通用户。提高进程资源的软硬限制主要通过函数 setrlimit()完成。

## 3．6 线程安全

线程安全，是指一个方法（method）可以在多线程的环境下安全有效的访问进程级的数据（这些数据是与其他线程共享的）。线程安全的核心概念就是同步，它保证多个线程：

- 同时开始执行，并行运行
- 不同时访问相同的对象实例
- 不同时执行同一段代码

使用上面的方法处理好全局变量问题之后，线程安全问题实际上只集中于程序使用的各种库函数。所以需要查出程序的所有线程不安全的库函数，不过，glibc 库在缺

省编译配置下是线程安全的[8]，可以省去不少的麻烦。

## 3．7 动态内存回收

当线程异常退出时，动态分配出去的内存必须回收，否则将造成内存泄漏。解决的方法是建立全局的动态内存回收链表，保存各个线程分配的动态内存的指针。当线程异常退出后，由主线程释放其动态内存。

## 4．PostgreSQL 短锁等待队列唤醒标志的设计缺陷及解决

## 4．1 短锁等待队列唤醒标志的功用分析

PostgreSQL 中短锁的基本作用是为多个进程并发访问（改造为多线程之后，应该为多个线程并发访问）共享内存中的数据结构提供互斥保护。每个短锁有一个等待队列，如果线程无法获得短锁，则加入该短锁的等待队列。

唤醒标志是一个布尔变量，主要用来决定是否释放短锁等待队列中的线程（每次只释放等待队列头部的一个或几个线程）。目前判定释放等待队列的条件是：短锁无人占用而且唤醒标志为真。

唤醒标志在下面两种情况下发生变化：

- 有线程进入等待队列，则将唤醒标志置为真。如果一直没有新的线程进入等待队列，则唤醒标志一直不变，保持假值。
- 释放等待队列头部的一个或多个线程后，将唤醒标志置为假。

## 4．2 缺陷分析

唤醒标志的使用有可能造成无限期等待或者死锁。因为：线程在队列中等待，如果其他线程每次都能顺利地获取短锁，不进入等待队列，那么唤醒标志将一直为假，线程将一直等待下去不被释放，如果线程持有锁，那么其他线程将一直等待线程释放该锁。另外一方面，PostgreSQL 的死锁检测机制并没有将短锁纳入检测范围，如果等

待短锁的线程事实上构成了等待图中环路的一条边，那么死锁检测机制将永远检测不到这个死锁。

在实际的调试过程中，也发现了这种现象。启动 250 个连接访问数据库，warehouse 的值设为 1（这样更容易产生冲突），运行 1~2 个小时以后，死锁发生，使用 gdb 调试工具查看各个线程的当前运行状况，大量的线程都在直接或间接地等待某个事务的提交，事务的执行者正在等待一个短锁而无人唤醒它。

将唤醒标志废弃不用以后，无限期等待和死锁不再发生。

## 4．3 解决方法

同时引入下面的解决方法：

- 废弃唤醒标志。释放短锁时，只要短锁已经无人占用，则直接唤醒一个或者多个线程，无需判定唤醒标志。
- 让等待线程自己唤醒自己。线程进入短锁等待队列后，设置闹钟，超过指定的时间后，如果短锁无人占用，则自己醒来，去获取短锁。

目前，在多线程版本中，同时采用这两种方法，能够长时间稳定地运行。

## 5．性能测试与比较

我们以上述方案为基础，完成了对 PostgreSQL 源码的多线程改造，目前，我们改造的多线程 PostgreSQL 已经能够在 linux 上稳定运行。

我们采用的测试机器的配置是: CPU 为 P4，主频 2G；内存 1G;硬盘 40G。

线程改造前的 TPCC 测试结果（测试的数据量为 20warehouse）如图 5(a)。线程改造后的 TPCC 测试结果（测试的数据量为 20warehouse）如图 5(b)。

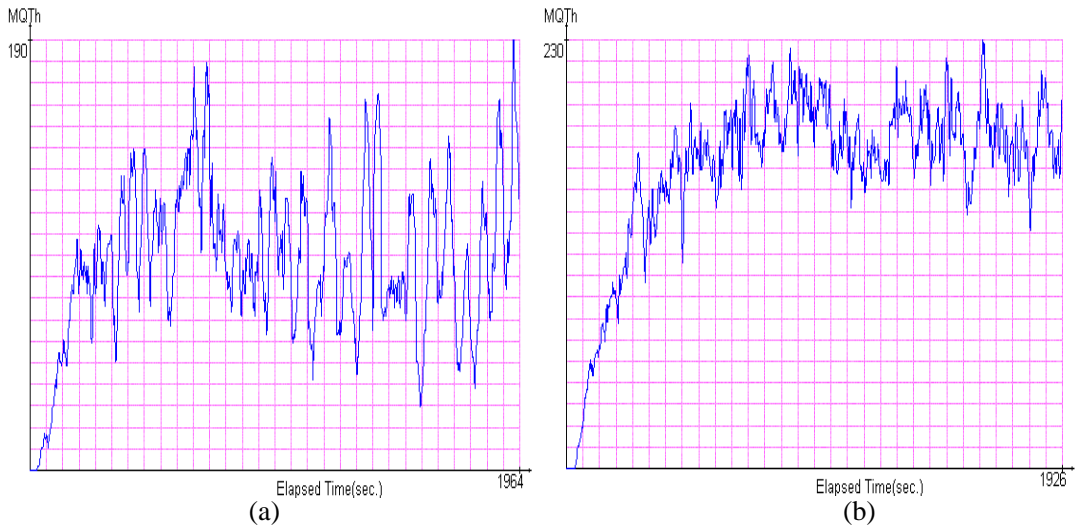

图 5　PostgreSQL 在线程改造前后的 TPCC 测试结果

## 6．小结

本文主要阐述一个改造 PostgreSQL 数据库管理系统为多线程结构的完整的实际方案。方案的主要指导思想是：以最少的代码改动量实现多线程改造，减少改造带来的不稳定影响。体现在三个方面：

- 代码比较独立的部分仍然保持进程结构，如：pgstat 父子进程；
- 尽量采用模拟机制，将代码改动限制在底层，而上层代码不变。如：对共享内存、信号量的处理；
- 全局变量采用全局数据结构与宏相结合的方法，避免代码改动。

文中对线程改造中的其他疑难问题也作了比较深入的探讨，如：信号机制的处理、PostgreSQL 短锁等待队列唤醒标志的设计缺陷及解决、进程资源限制、动态内存回收等等。

依据方案对 PostgreSQL 进行改造，得到了一个基本稳定且性能有所提高的 DBMS 内核。但是，改造工作仍然需要进一步完善和提高，以充分挖掘多线程的优势，提高

性能。如有兴趣请阅读本人的其他论文[10-49].

# References

1. Robert D. Blumofe, "Executing Multithreaded Programs Efficiently", Ph.D. thesis, Department of Electrical and Computer Science, Massachusetts Institute of Technology, September 1995.
2. C. Schmidtmann, M. Tao, and S. Watt. Design and implementation of a multithread Xlib. In USENIX Association. Proceedings of the Winter 1993 USENIX Conference. San Diego, CA, USA. USENIX, pages 193--203, 25-29 Jan. 1993.
3. R. D. Blumofe, C. F. Joerg, B. C. Kuszmaul, C. E. Leiserson, K. H. Randall, and Y. Zhou. Cilk: An Efficient Multithreaded Runtime System. In Proceedings of the 5th Symposium on Principles and Practice of Parallel Programming, 1995.
4. J. Lo, L. Barroso, S. Eggers, K. Gharachorloo, H. Levy, and S. Parekh, An analysis of database workload performance on simultaneous multithreaded processors, in "Proceedings of the Twenty-Fifth International Symposium on Computer Architecture", pp. 39--50, 1998.
5. W.Richard Stevens, Advanced Programming in the UNIX Environment, Addison Wesley Publishing Company, 1992.
6. Lewis and D. J. Berg, Multithreaded Programming with Pthreads, Prentice Hall, 1998.
7. K. A. Robins and S. Robins, Practical UNIX Programming: A Guide to Concurrency and Multithreading, Prentice Hall, 1996.
8. http://www.gnu.org/software/libc/FAQ.html
9. http://www.postgresql.org/docs/faq-english.html
10. Zhiyong Shan, Tanzirul Azim, Iulian Neamtiu. Finding Resume and Restart Errors in Android Applications. ACM Conference on Object-Oriented Programming, Systems, Languages & Applications (OOPSLA'16), November 2016.
11. Zhiyong Shan, I. Neamtiu, Z. Qian and D. Torrieri, "Proactive restart as cyber maneuver for Android," Military Communications Conference, MILCOM 2015 - 2015 IEEE, Tampa, FL, 2015, pp. 19-24.
12. Jin, Xinxin, Soyeon Park, Tianwei Sheng, Rishan Chen, Zhiyong Shan, and Yuanyuan Zhou. "FTXen: Making hypervisor resilient to hardware faults on relaxed cores." In 2015 IEEE HPCA'15, pp. 451-462. IEEE, 2015.
13. Zhiyong Shan, Xin Wang, Tzi-cker Chiueh: Shuttle: Facilitating Inter-Application Interactions for OS-Level Virtualization. IEEE Trans. Computers 63(5): 1220-1233 (2014)
14. Zhiyong Shan, Xin Wang: Growing Grapes in Your Computer to Defend Against Malware. IEEE Transactions on Information Forensics and Security 9(2): 196-207 (2014)
15. Zhiyong Shan, Xin Wang, Tzi-cker Chiueh: Malware Clearance for Secure Commitment of OS-Level Virtual Machines. IEEE Transactions on Dependable and Secure Computing. 10(2): 70-83 (2013)
16. Zhiyong Shan, Xin Wang, Tzi-cker Chiueh: Enforcing Mandatory Access Control in Commodity OS to Disable Malware. IEEE Transactions on Dependable and Secure Computing 9(4): 541-555 (2012)
17. Zhiyong Shan, Xin Wang, Tzi-cker Chiueh, Xiaofeng Meng: Facilitating inter-application interactions for OS-level virtualization. In Proceedings of the 8th ACM Annual International Conference on Virtual Execution Environments (VEE'12), 75-86
18. Zhiyong Shan, Xin Wang, Tzi-cker Chiueh, and Xiaofeng Meng. "Safe side effects commitment for OS-level virtualization." In Proceedings of the 8th ACM international conference on Autonomic computing (ICAC'11), pp. 111-120. ACM, 2011.
19. Zhiyong Shan, Xin Wang, and Tzi-cker Chiueh. 2011. Tracer: enforcing mandatory access control in commodity OS with the support of light-weight intrusion detection and tracing. In Proceedings of the 6th ACM Symposium on Information, Computer and Communications Security (ASIACCS '11). ACM, New York, NY, USA, 135-144. (full paper acceptance rate 16%)
20. Shan, Zhiyong, Tzi-cker Chiueh, and Xin Wang. "Virtualizing system and ordinary services in Windows-based OS-level virtual machines." In Proceedings of the 2011 ACM Symposium on Applied Computing, pp. 579-583. ACM, 2011.
21. Shan, Zhiyong, Yang Yu, and Tzi-cker Chiueh. "Confining windows inter-process communications for OS-level virtual machine." In Proceedings of the 1st EuroSys Workshop on Virtualization Technology for Dependable Systems, pp. 30-35. ACM, 2009.
22. Shan, Zhiyong. "Compatible and Usable Mandatory Access Control for Good-enough OS Security." In Electronic Commerce and Security, 2009. ISECS'09. Second International Symposium on, vol. 1, pp. 246-250. IEEE, 2009.
23. Xiao Li, Wenchang Shi, Zhaohui Liang, Bin Liang, Zhiyong Shan. Operating System Mechanisms for TPM-Based Lifetime Measurement of Process Integrity. Proceedings of the IEEE 6th International Conference on Mobile Adhoc and Sensor Systems (MASS 2009), Oct., 2009, Macau SAR, P.R.China, IEEE Computer Society. pp. 783--789.
24. Xiao Li, Wenchang Shi, Zhaohui Liang, Bin Liang, Zhiyong Shan. Design of an Architecture for Process Runtime Integrity Measurement. Microelectronics & Computer, Vol.26, No.9, Sep 2009:183~186. (in Chinese)
25. Zhiyong Shan, Wenchang Shi. "STBAC: A New Access Control Model for Operating System". Journal of Computer Research and Development, Vol.45, No.5, 2008: 758~764.(in Chinese)
26. Liang Wang, Yuepeng Li, Zhiyong Shan, Xiaoping Yang. Dependency Graph based Intrusion Detection. National Computer Security Conference, 2008. (in Chinese)
27. Zhiyong Shan, Wenchang Shi. "An Access Control Model for Enhancing Survivability". Computer Engineering and Applications, 2008.12. (in Chinese)
28. Shi Wen Chang, Shan Zhi-Yong. "A Method for Studying Fine Grained Trust Chain on Operating System", Computer Science, Vol.35, No.9, 2008, 35(9):1-4. (in Chinese)
29. Liang B, Liu H, Shi W, Shan Z. Automatic detection of integer sign vulnerabilities. In International Conference on Information and Automation, ICIA 2008. (pp. 1204-1209). IEEE.
30. Zhiyong Shan, Qiuyue Wang, Xiaofeng Meng. "An OS Security Protection Model for Defeating Attacks from Network", the Third International Conference on Information Systems Security (ICISS 2007), 25-36.
31. Zhiyong Shan, "A Security Administration Framework for Security OS Following CC", Computer Engineering, 2007.5, 33(09):151-163. (in Chinese)
32. Shan Zhiyong, "Research on Framework for Multi-policy", Computer Engineering, 2007.5, 33(09):148-160. (in Chinese)
33. Zhiyong Shan, Shi Wenchang, Liao Bin. "Research on the Hierarchical and Distributed Network Security Management System". Computer Engineering and Applications, 2007.3, 43(2):20-24. (in Chinese)
34. Zhiyong Shan, "An Architecture for the Hierarchical and Distributed Network Security Management System", Computer Engineering and Designing, 2007.7, 28(14):3316-3320. (in Chinese)
35. Shan Zhi Yong, Sun Yu Fang, "Study and Implementation of Double-Levels-Cache GFAC", Chinese Journal of Computers, Nov, 2004, 27(11):1576-1584. (in Chinese)
36. Zhiyong Shan, Yufang Sun, "An Operating System Oriented RBAC Model and Its Implementation", Journal of Computer Research and Development, Feb, 2004, 41(2):287-298. (in Chinese)
37. Zhiyong Shan, Yufang Sun, "A Study of Extending Generalized Framework for Access Control", Journal of Computer Research and Development, Feb, 2003, 40(2):228-234. (in Chinese)
38. Shan Zhi Yong, Sun Yu Fang, "A Study of Generalized Environment-Adaptable Multi-Policies Supporting Framework", Journal of Computer Research and Development, Feb, 2003, 40(2):235-244. (in Chinese)
39. Shan Zhiyong, Research on the Framework for Multi-Policies and Practice in Secure Operation System. Phd Thesis, Institute of Software, Chinese Academy of Science 2003. (in Chinese)
40. Shan Zhi Yong, Sun Yu Fang, "A Study of Security Attributes Immediate Revocation in Secure OS", Journal of Computer Research and Development, Dec, 2002, 39(12):1681-1688. (in Chinese)
41. Shi Wen Chang, Sun Yu Fang, Liang Hong Liang, Zhang Xiang Feng, Zhao Qing Song, Shan Zhi Yong. Design and Implementation of Secure Linux Kernel Security Functions. Journal of Computer Research and Development, 2001, 38(10), 1255-1261.
42. Zhiyong Shan, Tzi-cker Chiueh, Xin Wang. Duplication of Windows Services. CoRR, 2016.
43. Zhiyong Shan. Suspicious-Taint-Based Access Control for Protecting OS from Network Attacks. CoRR, 2016.
44. Zhiyong Shan, Bin Liao. Design and Implementation of A Network Security Management System. CoRR, 2016.
45. Zhiyong Shan. A Study on Altering PostgreSQL From Multi-Processes Structure to Multi-Threads Structure. Technical Report, 2014.
46. Zhiyong Shan. Implementing RBAC model in An Operating System Kernel. Technical Report, 2015.
47. Zhiyong Shan. A Hierarchical and Distributed System for Handling Urgent Security Events. Technical Report, 2014.
48. Zhiyong Shan. An Review On Thirty Years of Study On Secure Database and It`s Architectures. Technical Report, 2014.
49. Zhiyong Shan. An Review on Behavior-Based Malware Detection Technologies on Operating System. Technical Report, 2014.